\documentclass[11pt]{article}

\usepackage[normalem]{ulem}
\usepackage[swedish,english]{babel}

\usepackage{color}
\usepackage{xcolor}
\usepackage{amssymb}
\usepackage{amsmath}

\usepackage{ae}
\usepackage{slashed}
\usepackage{graphics}
\usepackage{graphicx}
\usepackage{epstopdf}
\usepackage{url}
\usepackage{hyperref}

\usepackage{cite}

\usepackage{times}
\usepackage{xfrac}

\usepackage{fancyhdr}
\usepackage{enumerate}
\def\be{\begin{equation}}
\def\ee{\end{equation}}

\usepackage{tikz}
\begin{document}
\selectlanguage{english}
\frenchspacing
\pagenumbering{roman}
\begin{center}
\null\vspace{\stretch{1}}

{ \Large {\bf 
Replica wormhole and island incompatibility\\ with monogamy of entanglement
}}
\\
\vspace{1cm}
Anna Karlsson$^{1,2}$
\vspace{1cm}

{\small $^{1}${\it
Institute for Advanced Study, School of Natural Sciences\\
1 Einstein Drive, Princeton, NJ 08540, USA}}
\vspace{0.5cm}

{\small $^{2}${\it
Division of Subatomic, High Energy and Plasma Physics, Department of Physics, \\
Chalmers University of Technology, 412 96 Gothenburg, Sweden}}
\vspace{1.6cm}
\end{center}

\begin{abstract}
We argue that the `island conjecture' and the replica wormhole derivation of the Page curve break monogamy of entanglement through allowing black hole interior states to be non-classically correlated while also pairwise entangled with radiation states. The reason is that quantum degrees of freedom (present in any half of a Hawking pair) cannot all be identified with the environment at semi-classical pair production, and can only be fixed relative to a subsystem, as required for the Page curve, by correlations equivalent to entanglement --- regardless of what those correlations are attributed to. This implies that the recent gravity (replica wormhole) and holographic (island conjecture) derivations of the Page curve entail new physics not yet properly taken into account.
\end{abstract}

\vspace{\stretch{3}}
\thispagestyle{empty}
\newpage
\pagenumbering{arabic}

\section{Introduction}
The recent development regarding the black hole information paradox \cite{Hawking:1974sw,Hawking:1976ra} is that the Page curve \cite{Page:1993wv,Page:2013dx} can be derived using standard methods, without a requirement of new physics --- standard quantum mechanics and no new physics at the event horizon is sufficient. The procedure for how to obtain the Page curve for the entropy of Hawking radiation was first developed in a holographic setting \cite{Penington:2019npb,Almheiri:2019psf,Almheiri:2019hni} and then extended to a calculation directly in the gravity theory, using replica wormholes \cite{Penington:2019kki,Almheiri:2019qdq}. Further related investigations include \cite{Marolf:2020xie,Akers:2019nfi,Almheiri:2019yqk,Bousso:2019ykv,Almheiri:2019psy,Liu:2020gnp,Piroli:2020dlx,Balasubramanian:2020hfs,Verlinde:2020upt,Chen:2020wiq,Gautason:2020tmk,Giddings:2020yes,McNamara:2020uza,Sully:2020pza,Hartman:2020swn,Chen:2020uac,Dong:2020iod,Marolf:2020vsi,Bousso:2020kmy,Anous:2020lka,Dong:2020uxp}, and \cite{Almheiri:2020cfm} provides a review.

The replica wormhole derivation and the island conjecture reproduce the Page curve. However, we argue that a reproduction of the Page curve is unlikely to fit within standard physics. Under standard physics, black hole evaporation takes place through pair production near the horizon. In this scenario, monogamy of entanglement \cite{PhysRevA.61.052306} typically has been discussed in terms of the paradox in \cite{Almheiri:2012rt}, as commented on in \cite{Almheiri:2020cfm}. The currently suggested solution is that at late times, a new Hawking radiation mode does not need to be entangled with two separate systems (its interior partner and the previously emitted Hawking radiation) due to that the interior is more intricate than previously appreciated, so that the interior partner can be identified with some of the earlier Hawking radiation. The entanglement then is with one system instead of two. This solution does not address the issue we raise: how either part of a new Hawking pair is supposed to entangle with anything but each other. In semi-classical pair production, both parts are formed close to the horizon, and entangled with each other at that formation. For either part to become entangled or identified with anything else, it must first disentangle from its partner. If this is not the case, the emission is either not by pair production\footnote{A model where a Hawking radiation mode is entangled with the earlier radiation the instant it is formed would be indistinguishable from emission of the black hole interior. At the very least, every mode cannot be identified in that way.}, or monogamy of entanglement is broken. If the Hawking pair does disentangle, the process is different from that suggested in the island conjecture and the replica wormhole setup.
 
Our argument presents a conceptual sharpening of the argument in \cite{Mathur:2009hf} through limitations set by monogamy of entanglement for each Hawking pair. The concern is for the behaviour of the quantum degrees of freedom of each individual Hawking mode, and how that limits a reproduction of the Page curve. While independence of Hawking pair production from its surroundings is a truth with modification --- leading to the small corrections discussed in \cite{Mathur:2009hf} and ostensibly allowing for the replica wormhole setup --- the concept of monogamy of entanglement is an absolute. The Page curve feature of $S_\text{rad}(t)\not\geq \varepsilon t$ for some emission rate $\varepsilon >0$ requires that the added quantum degrees of freedom (presented by one half of the Hawking pair) become correlated with the rest of the Hawking radiation. A correlation specifying (or for the interior mode: `identifying') non-classical degrees of freedom by definition represents entanglement, regardless of what theoretical model the correlations are attributed to, and cannot manifest itself for \emph{either} part (exterior or interior mode) while it is still entangled with its partner --- not without violating monogamy of entanglement.

In the replica wormhole derivation and the island conjecture, it is easy to overlook the property of non-monogamy, since it is inferred by the definitions made --- of the state overlaps in the replica wormhole setup, and of the island conjecture formula, respectively --- and present despite the semi-classical nature of the final formulas. In addition, the replica wormhole derivation and the island conjecture rely on density matrix theory e.g. for the expression for the radiation entropy. Since entanglement is monogamous in density matrix theory, it is problematic to employ that theory in the presence of non-monogamous entanglement. Effectively, the two methods introduce new physics (non-monogamy, contrary to the present claim) and the treatment risks being inconsistent.

What happens in the two above mentioned approaches to obtaining the Page curve is the following. To introduce new behaviour around the Page time without introducing new physics, a new mechanism is defined which gives the desired effect, but breaks monogamy without that being recognized. The setting then includes entanglement within the separate subsystems of the interior and radiation in parallel with those same states being entangled in Hawking pairs. At the same time, density matrix theory is used to describe the physics. For compatibility with density matrix theory, the correlations inferred by replicas, or through higher dimensions in the manner of ER=EPR \cite{Einstein:1935rr,Einstein:1935tc,Maldacena:2001kr,Swingle:2009bg,VanRaamsdonk:2010pw,Hartman:2013qma,Maldacena:2013xja} intended in the holographic setting, need to be restricted to classical correlations when the states under consideration simultaneously are pairwise entangled. Otherwise, the methods (including the non-standard feature of ensemble averaging) need to be proven consistent, general and accurate with respect to the new physics (e.g. non-monogamy in the interior). We discuss the scenarios in more detail in \S\ref{s.cond}, and make a brief summary of the remaining solutions to the information paradox in \S\ref{s.outlook}.

\section{Restrictions under standard physics}\label{s.cond}
In the general setting of the black hole information paradox, two central assumptions are made,
\begin{itemize} 
\item[\emph{(i)}] quantum mechanics as described by density matrix theory, including e.g. unitarity (no information loss) and monogamy of entanglement,
\item[\emph{(ii)}] semi-classical gravity with nothing special at black hole event horizons.
\end{itemize}
This restricts black hole evaporation to take place through pair production close to the horizon, where a pure state separating into two pieces are formed. One part escapes to become part of the Hawking radiation while its partner becomes part of the black hole interior. Since the pair describes a pure state, its formation does not to add entropy to the total system, $\Delta S=0$. Each such pair is entangled, and the parts have quantum degrees of freedom. In the idealized case, the pair production is independent of properties pertaining to the black hole or anything outside it --- it just forms out of the vacuum. In reality, small corrections to the idealized case are allowed, since the pair is formed in an environment. This e.g. opens up for the replica wormhole setup, which presents a different type of correction from those ruled out in \cite{Mathur:2009hf}, but the physical restrictions are still more severe.

For the entropy of Hawking radiation to follow the Page curve, it is required that\footnote{We keep the argument general and independent of what is emitted. The only condition is that there is an emission of non-classical degrees of freedom formed independently of the environment, so that $\varepsilon>0$. $\varepsilon$ represents a rate of emission of the smallest non-classical, independent entropy contribution from one of the two modes in a Hawking pair. When the only tracked emission is of (independent) quantum degrees of freedom, whose contribution to the entropy cannot be reduced by classical correlations, one can argue that $\partial_t S_\text{rad}\geq\varepsilon$ is required. $S_\text{rad}(t)\geq \varepsilon t$ also holds for the model in \cite{Karlsson:2019vlf}.}
\be\label{eq.nS}
S_\text{rad}(t)\not\geq \varepsilon t\,,\quad \varepsilon >0\,,
\ee
at some point around or after the Page time, for eternal and evaporating black holes. At black hole evaporation, each state added to the Hawking radiation starts out entangled with an interior mode. The pair production is nearly independent (direct identification with other states is not accommodated). While a state in the Hawking radiation remains entangled with its interior partner, it cannot be entangled with any other state (by monogamy of entanglement); nor can its partner. Consequently, at that time the mode in the Hawking radiation can only be classically correlated with the rest of the Hawking radiation, and since classical correlations cannot reduce quantum degrees of freedom, it gives a non-reducible contribution to $S_\text{rad}$ in terms of its non-classical constituents,
\be\label{eq.gS}
\Delta S_\text{rad, \text{ quant}}>0\,,
\ee
contrary to \eqref{eq.nS}. This argument is a conceptual rephrasing of the $S_{bE}-S_E\geq S_b$ of \cite{Mathur:2009hf}, where $b$ is a newly added Hawking mode and $E$ is the prior Hawking radiation. However, the bound is not a technical limitation (by small deviations from the ideal state configuration) as was the focus of \cite{Mathur:2009hf}, but a restriction by monogamy of entanglement. Whatever corrections to Hawking pair production one considers, the pair produced modes must be entangled, not all of those entangled degrees of freedom can initially be identified with the environment (or the evaporation is not by pair production), and monogamy of entanglement limits further evolution of the modes.

For the effect of \eqref{eq.nS}, radiation states must entangle with each other. With monogamy of entanglement, the independent, pair produced modes must first disentangle, i.e., given two different states $A$ and $B$ that are not entangled, their correlations to other states must be reduced to be classical before $A$ can entangle with $B$. There must be an intermediary state that only is governed by classical correlations, i.e. correlations that cannot give rise to entanglement. Consequently, under pair production and monogamy of entanglement, \eqref{eq.nS} requires classically correlated radiation modes to \emph{spontaneously} entangle with each other, since guidance is restricted to classical correlations.
 
\subsection{Remarks on the replica wormhole setup}\label{s.rw}
Moving on to the entanglement present in the replica wormhole derivation, the setup is the most straightforward in \cite{Penington:2019kki}, where the details are explicit. Their toy model of an evaporating black hole displays the basic principles. The state of the full system is
\be\label{eq.pure}
|\Psi\rangle=\frac{1}{\sqrt{k}}\sum_{i=1}^k|\psi_i\rangle_B|i\rangle_R\,,
\ee
where $|\psi_i\rangle_B$ is a state of the black hole and $|i\rangle_R$ is a state of the radiation system. The gravitational path integral specifies that these states be characterized by
\be
\langle\psi_i|\psi_j\rangle=\delta_{ij}\,,\quad |\langle\psi_i|\psi_j\rangle|^2=\delta_{ij}+Z_2/Z_1^2\,.
\ee
Due to the incompatibility, the result is reinterpreted as an ensemble average, i.e. applicable to $\overline{\langle\psi_i|\psi_j\rangle}$ and $\overline{|\langle\psi_i|\psi_j\rangle|^2}$ instead of the non-averaged expressions above. With $Z_i$ a specific gravitational path integral, the average is over microscopic quantities $r_{ij}$ and
\be\label{eq.rwp}
\langle\psi_i|\psi_j\rangle=\delta_{ij}+e^{-S_0/2}r_{ij}\,,\quad \overline{r_{ij}}=0\,,\quad \overline{|r_{ij}|^2}=1\,,
\ee
with the normalization implicit. This is concluded to make the equations compatible, and the rest of the argument builds on this setup of $|\psi_i\rangle$ and $|i\rangle$; the replica wormholes specify the overlap of \eqref{eq.rwp} and the entropy of the radiation. The term $e^{-S_0/2}$ introduces a new scale at the Page time, which reproduces the Page curve instead of Hawking's result. For example, the second R\'enyi entropy is
\be\label{eq.S2}
S_2(\rho_R)=-\log\operatorname{tr}(\rho_R^2)\,,\qquad \operatorname{tr}(\rho_R^2)=\frac{1}{k^2}\sum_{i,j=1}^k|\langle\psi_i|\psi_j\rangle|^2\quad\stackrel{\eqref{eq.rwp}}{\sim}\quad k^{-1}+e^{-S_0}\,,
\ee
where the expression to the far right is schematic and $k^{-1} \ll e^{-S_0}$ at late times. This additional effect carries over to an analytic continuation in $n$ (number of replicas) and specifies an `island' contribution to the von Neumann entropy through $S(\rho_R)=-\lim_{n\rightarrow1} \frac{1}{n-1}\log\operatorname{tr}(\rho_R^n)$. \cite{Penington:2019kki}

What is important here is not only that \eqref{eq.rwp} can reproduce the Page curve, but also if \eqref{eq.rwp} is consistent with the physical assumptions made, i.e. those of \emph{(i-ii)} above. Based on the reasoning between equations \eqref{eq.nS} and \eqref{eq.gS} a contradiction occurs, but where? The replica wormhole ansatz has two key elements, \eqref{eq.pure} and \eqref{eq.rwp}. \eqref{eq.pure} infers that the pair entanglement remains throughout the evaporation process. \eqref{eq.rwp} (in combination with \eqref{eq.pure} and density matrix properties) caps the entropy of the radiation in a way equivalent to entanglement between quantum degrees of freedom in the radiation subsystem. As noted above, the relevant feature is not what the correlations are attributed to, but what they do from the point of view of the degrees of freedom --- i.e. they fix quantum degrees of freedom relative to one another, as only non-classical correlations can do. Since \eqref{eq.nS} cannot arise from pair production under \eqref{eq.pure} and the assumptions of \emph{(i-ii)}, the definition \eqref{eq.rwp} either assumes non-monogamy of entanglement between the pair and its environment directly at the pair production, or a further evolution of the interior states to the same effect.

The term proportional to $e^{-S_0/2}$ in \eqref{eq.rwp} is what enables a reproduction of the Page curve. However, the fact that it through \eqref{eq.pure} and the analytic continuation in $n$ produces the result of \eqref{eq.nS}, incompatible with the working assumptions, infers that \eqref{eq.rwp} introduces a presence of entanglement between Hawking modes \emph{within the same subsystem} ($R$ or $B$) at the same time as each mode is entangled with its partner in the other subsystem. The combination of \eqref{eq.rwp} and \eqref{eq.pure} breaks monogamy of entanglement in the limit $n\rightarrow1$. It does not matter that the correction in \eqref{eq.rwp} is small.

\subsection{Remarks on the island conjecture}
The concern regarding monogamy of entanglement also extends to the holographic calculation and the island conjecture. The identification of a new quantum extremal surface that could explain the Page curve \cite{Penington:2019npb,Almheiri:2019psf} led to a conjecture of a new entropy rule for the Hawking radiation \cite{Almheiri:2019hni}, summarized in \cite{Almheiri:2019qdq} as
\be
S(\rho_R)=\text{ext}_Q\left[\frac{\text{Area}(Q)}{4G_N}+S(\tilde\rho_{R\cup I})\right]\,,
\ee
with $R$ the region the radiation is in, $I$ the island, $Q$ the quantum extremal surface and $\tilde\rho$ the density matrix in the semi-classical theory. This entropy rule includes a term capping the entropy through the incorporation of an `island', without which the radiation entropy follows Hawking's calculation \cite{Hawking:1976ra} instead of the Page curve \cite{Almheiri:2019hni}. The island is disconnected from the entanglement wedge of the radiation, but in \cite{Almheiri:2019hni} the two are considered to be connected by an extra dimension, possible to view `as a realization of the ER=EPR idea'. This type of additional connection is problematic in terms of entanglement since the interior and radiation already are entangled through (nearly independent) pair production, and monogamy of entanglement states that additional correlations between such states are restricted to classical correlations, unlikely to fit within the type of connection intended in \cite{Almheiri:2019hni}. From a practical perspective, the problem is how $S_\text{rad}(t)\not\geq \varepsilon t$ should occur from pair production while each pair remains entangled (limiting further correlations for \emph{both} parts).

\section{Outlook}\label{s.outlook}
Recall the last sentence before \S\ref{s.rw}. Since the Page curve is not the product of a random process (Hawking's result is only corrected around and after the Page time), from the point of view of how to reproduce it, we either must have
\begin{itemize}
\item[a)] a way to detail how the Page curve spontaneously arises in the radiation subsystem, based on initial conditions that are limited to classical correlations,
\item[b)] the entropy of Hawking radiation does not follow the Page curve due to a presence of some remnant (a separate system, e.g. a baby universe, not accounted for in the derivation of the Page curve) and instead has\footnote{Assuming the radiation escapes, and is not reflected to fall back into the black hole.} $S_\text{rad}(t)\geq \varepsilon t$,
\item[c)] a change in quantum mechanics (altering the assumptions under which the Page curve was derived), or of the dynamics near the black hole event horizon.
\end{itemize}
Of these, a) is improbable, b) is an ad hoc solution and c) would alter the assumptions \emph{(i-ii)} and mean important conceptual changes in physics (where new physics near the horizon perhaps is the most uncontroversial scenario). With respect to the black hole information paradox, b) represents new black hole physics. A disentangling of Hawking radiation modes (a) is also considered incompatible with an uneventful event horizon. Further evolution after some propagation of the modes might be a different matter though, or require a constraint. Such interaction with an environment is commonplace for e.g. spin 1/2 particles, regardless of if they originate in pair production or not.

One alternative may be a variation on c): modifying the description of quantum mechanics (monogamy or locality) without altering the observable physics or the Page curve result. This scenario would require previously unrecognised features of the vacuum. Since the correlations to pair produced modes that give rise to the Page curve cannot be specific to pair production at the event horizon, any pair produced modes $A,B$ must at their formation be maximally entangled with some vacuum modes $C_i$ in a different region, even in the absence of an event horizon. This \emph{might} allow for a so-called $A=R_B$ scenario (where the interior partner $A$ to a radiation mode $B$ for one reason or another is `identified' with the early radiation $R_B$) to be modified to not introduce new physics (at the event horizon). A suggested scenario exists in \cite{Raju:2016vsu,Raju:2018zpn} (building on \cite{Papadodimas:2012aq}), where the modes $C_i$ are conjectured to be dependent on the vacuum and entangled with $R_B$ through its presence in the vacuum, and to give rise to the Page curve due to properties (e.g. the entanglement structure) of the vacuum configuration. However, note that the modes $C_i$ would have to be impossible to probe (classically) by \emph{any} system within a finite time\footnote{The maximal entanglement between three (irreducible) modes (e.g. $A,B$ and $C_i$, or a mode in $R_B$) required for the discussed entropy contribution $\varepsilon\not>0$ is not allowed in a causal, semi-classical theory. Under those conditions, there is a contradiction between that measurement outcomes constitute classical information, and that measurement outcomes from three such modes are incompatible with classical theory. The probabilities of the outcomes $\not\in[0,1]$, e.g. as shown for spin 1/2 particles in Bell's theorem \cite{Bell:1964kc}. In the absence of an event horizon (restricting the degree to which information coexists), the only possible loophole is that one of the modes is impossible to probe.}, and that the correlation process (between $A,B,C_i$ and $R_B$) must not introduce entanglement between $A,B$ and $R_B$ except through evolution after the pair production, provided that the pair produced modes are separated by an event horizon and that the evolution does not infer new physics at the event horizon. Consistency of these requirements remains to be shown. Typically, modes disentangle before entangling with new modes; the $C_i$ would have to work differently. Either way, such correlations would not be described by (local) density matrix theory, due to its property of monogamy of entanglement.

The main point above is that even an island cannot accommodate the presence of a Page curve for the entropy of Hawking radiation under the assumptions \emph{(i-ii)} as advocated in the island conjecture and the replica wormhole derivation. While a semi-classical theory near and outside the event horizon might be satisfactory, non-semi-classical behaviour somewhere in the interior would have an impact on the validity of the formalism employed. When introducing wormholes (of any kind) into a theory, it is relevant to keep in mind that they correspond to classically non-local identifications/correlations and risk being equivalent to entanglement. In general, a balance is needed between modelling theory on a desired physical scenario vs analyses of what the physics allows for.

\section*{Acknowledgements}
This work is supported by the Swedish Research Council grant 2017-00328. We are grateful to the hospitality of the KITP, Santa Barbara, where this work was initiated. This research was supported in part by the National Science Foundation under Grant No. NSF PHY-1748958.

\providecommand{\href}[2]{#2}\begingroup\raggedright\endgroup
\end{document}